\documentclass{cargese}\usepackage{graphicx,amssymb,epsfig}

\let\footnote\savefootnote
\let\footnotetext\savefootnotetext 
\let\footnoterule\savefootnoterule 
\normallatexbib

\def\bX{\mathbf X}

\def\IR{\mathbb{R}}

\def\beq{\begin{equation}}
\def\eeq{\end{equation}}

\newcommand{\fref}[1]{Figure~\ref{#1}}
\def\eref#1{(\ref{#1})}

\begin{document}

\articletitle[Chaotic Cascades for D-branes on Singularities]
{Chaotic Cascades for D-branes on Singularities}


\author{Sebasti\'{a}n Franco$^1$, Yang-Hui He$^2$, Christopher Herzog$^3$, Johannes Walcher$^4$}



\affil{1. \parbox[t]{6in}{Center for Theoretical Physics,
Massachusetts Institute of Technology,
Cambridge, MA 02139, USA}\\
2. \parbox[t]{6in}{Dept.~of Physics and Math/Physics RG,
Univ.~ of Pennsylvania,
Philadelphia, PA 19104, USA}\\
3. \parbox[t]{6in}{Kavli Institute for Theoretical Physics,
University of California,
Santa Barbara, CA 93106, USA} \\
4. \parbox[t]{6in}{Institute for Advanced Study,
Princeton, NJ 08540, USA}
}    

\email{sfranco@mit.edu,yanghe@physics.upenn.edu,herzog@kitp.ucsb.edu,walcher@ias.edu}


\begin{abstract}
We briefly review our work on
the cascading renormalization group flows for gauge theories on D-branes 
probing Calabi-Yau singularities. 
Such RG flows are sometimes chaotic and 
exhibit duality walls. We construct supergravity solutions dual to logarithmic flows 
for these theories.  We make new observations about a surface of conformal
theories and more complicated supergravity solutions.
\end{abstract}


\section{Introduction}

Extending the revolutionary AdS/CFT correspondence
\cite{Maldacena:1997re} beyond the original relation between 
$\mathcal{N}=4$ SYM on $N$ D3-branes and
Type IIB supergravity (sugra) in $AdS_5 \times S^5$ with $N$ units of RR
5-form flux on the $S^5$ is important to understanding
realistic strongly coupled field theories such as QCD. 

Two standard extensions have been (1) {\em reducing the SUSY to
$\mathcal{N}=1$} by placing the D3-branes transverse to a Calabi-Yau
singularity (the dual sugra background becomes 
$AdS_5 \times X^5$, where $X^5$ is some non-spherical horizon);
and (2) {\em breaking conformal invariance and inducing an RG flow}, 
by introducing fractional branes, i.e., D5-branes
wrapped over collapsing 2-cycles of the singularity 
(in the sugra dual, 3-form fluxes are turned on).
A fascinating type of RG flow is the {\em duality cascade}:
Seiberg duality is used to switch to an alternative
description whenever infinite coupling is reached. This idea was introduced
in \cite{Klebanov:2000hb} for the gauge theory on D-branes probing the conifold.

\section{Cascades in coupling space}

\begin{figure}
\[\begin{array}{ccc}
\hspace{-1in}
\begin{array}{cc}
\epsfxsize = 4cm \epsfbox{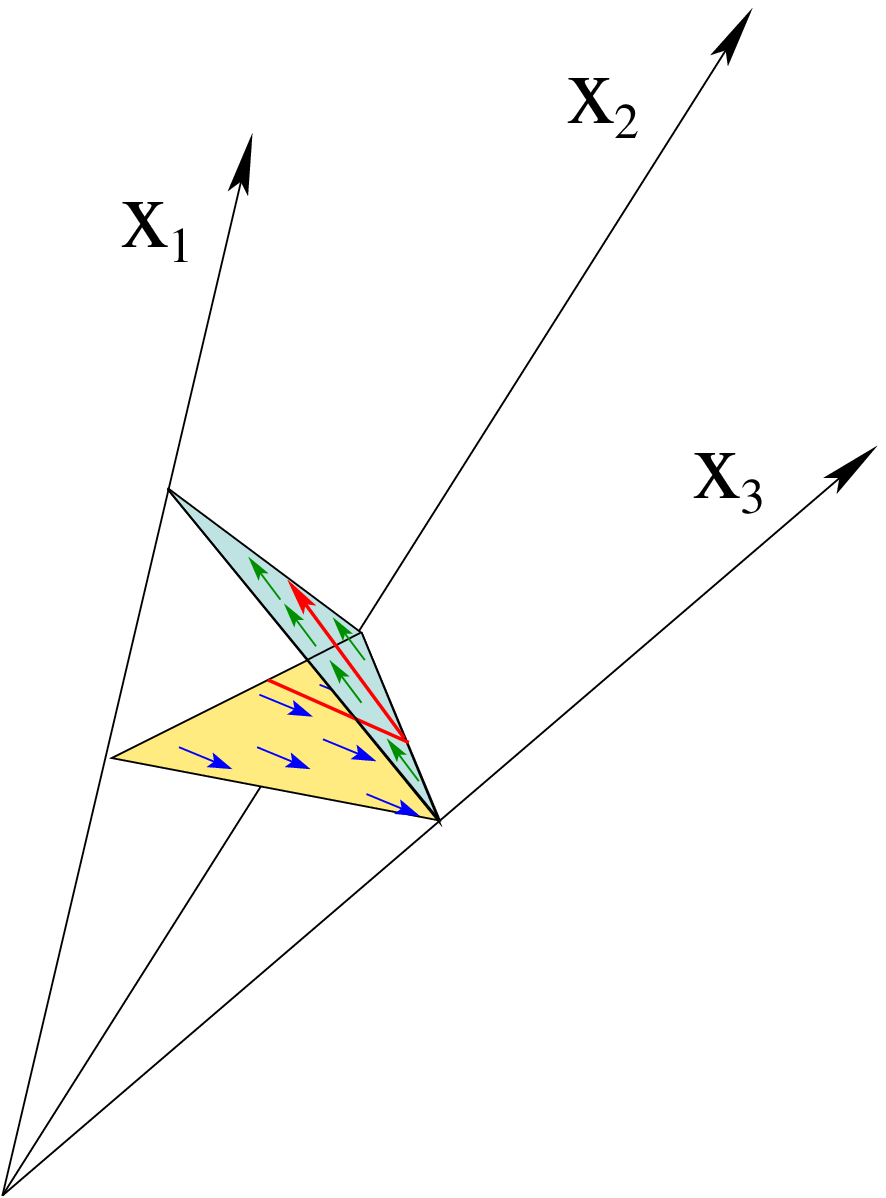}&(A)\\
\epsfxsize = 5cm \epsfbox{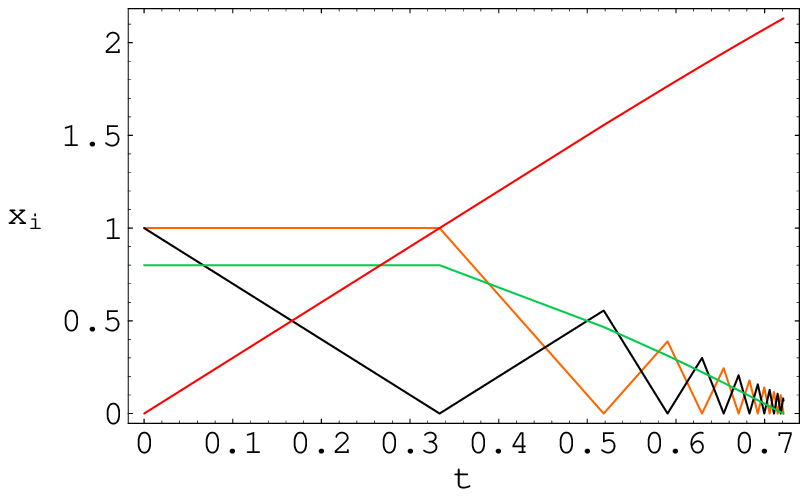}&(B)
\end{array}
&~&
\begin{array}{cc}
\epsfxsize = 8cm \epsfbox{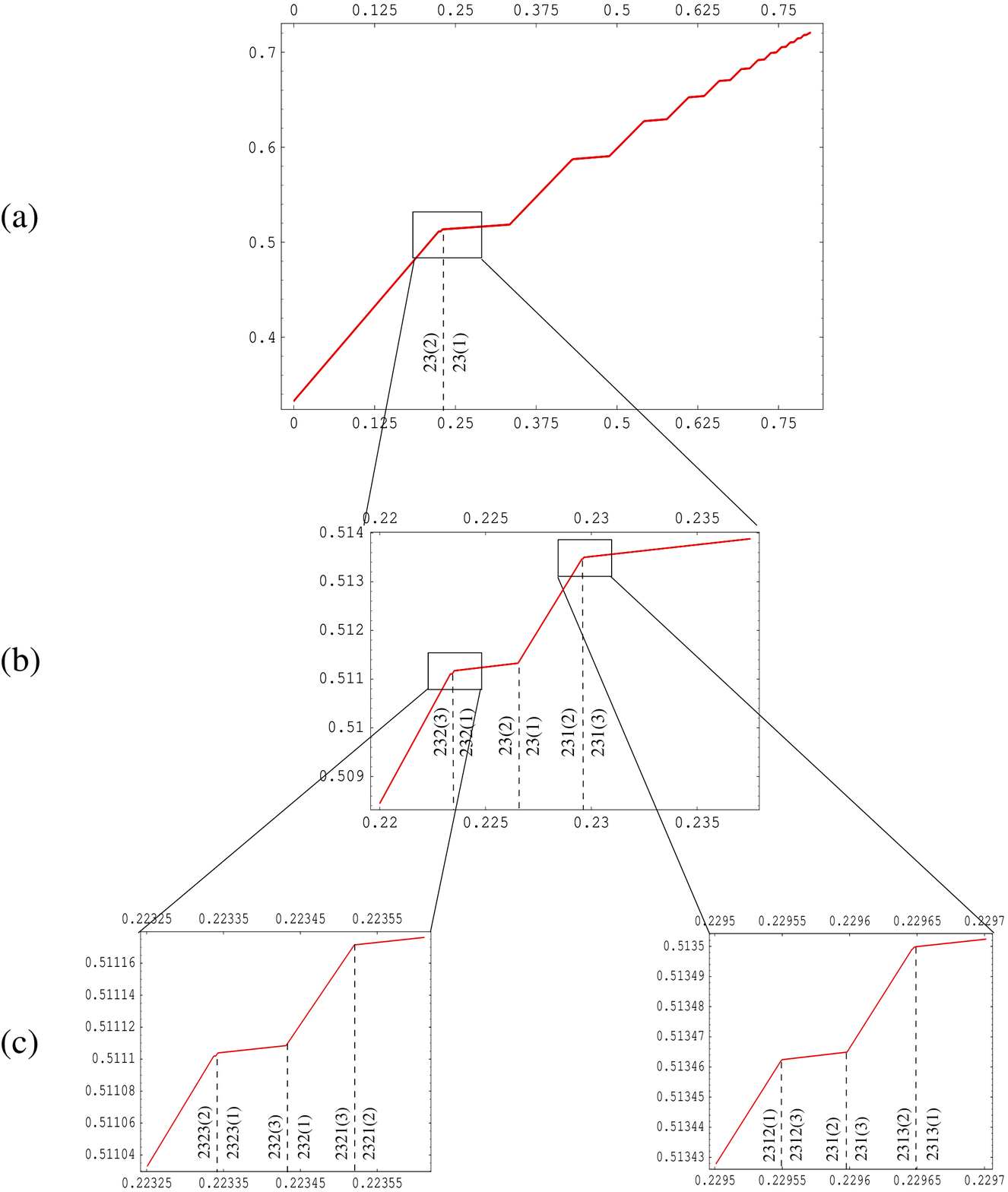}&(C)
\end{array}
\end{array}\]
\caption{}\label{f:fig}
\end{figure}

There is an interesting way to look at cascading RG flows. 
In a gauge theory described by a quiver with $k$ gauge groups, 
the inverse squared couplings $x_i \equiv 1/g_i^2$ are positive and 
define a $k$-dimensional cone $(\IR_+)^k$. 
Inside this cone,
the RG flow generates a trajectory dictated by the beta functions
and satisfying $\sum_i^k r^i/g_i^2 = \mbox{constant}$.
Each step between dualizations then corresponds to a straight line in
the simplex defined by the intersection between this hyperplane and
the $(\IR_+)^k$ cone. We show such a trajectory in \fref{f:fig}(A).
Now, each wall of the cone corresponds to one of the gauge couplings
going to infinity. 
Therefore, whenever one of them is reached, we switch to a Seiberg
dual theory at weak coupling.
There will be then a different simplex associated to the dual theory. 
The entire cascade corresponds to a flow in the space of glued simplices. 
From this perspective, 
which resembles a billiard bouncing in coupling space, 
one foresees that cascading RG flows 
will exhibit chaotic behavior.

\section{Duality Walls and Fractals}

After introducing the notion of a duality cascade, it is
natural to wonder whether some  
supersymmetric extension of the Standard Model, such as the MSSM, 
can sit at the IR endpoint of  a cascade. 
This question was posed by Matthew Strassler \cite{Strassler}. 
Generically, while trying to 
reconstruct such a RG flow, one encounters a UV accumulation point
beyond which Seiberg duality cannot proceed. 
This phenomenon is dubbed a {\em duality wall} and has been
constructed for
gauge theories engineered with D-branes on singularities \cite{Franco:2003ja}. 
\fref{f:fig}(B) shows the behavior of couplings for a cascade with a
duality wall for the theory on D-branes over a complex cone
over the Zeroth-Hirzebruch surface $F_0$ 
\footnote{We refer the reader to \cite{Franco:2003ja,Franco:2004jz}
for a detailed description of the associated quiver theory.}.

Postponing the question of a possible UV completion of duality walls,
we can study the dependence of its position on initial couplings.
Illustrating with $F_0$, the result is remarkable and is presented
in \fref{f:fig}(C). 
The curve is a fractal, with concave and convex cusps. 
Whenever we zoom in on a convex cusp, an infinite, self-similar structure
of more cusps emerges.

 One subtle point which was not emphasized in \cite{Franco:2004jz}
involves the existence in coupling space of a codimension two surface
of conformal theories for $F_0$ and the other del Pezzo quiver gauge theories.
If the number of gauge couplings is $n+2$, then a naive counting of the 
linearly independent $\beta$-functions constrains only two combinations
of gauge couplings when the theory is conformal, leaving an $n$-dimensional
surface of conformal theories.  This $n$-dimensional surface is parametrized
on the gravity side by the dilaton and the integral of the NSNS $B_2$ form
through $n-1$ independent 2-cycles.

The existence of this codimension two surface may well affect the existence
and behavior of the duality wall for $F_0$.  In \cite{Franco:2003ja}, it was
assumed that a generic choice of initial couplings would lie on the conformal
surface.  However, if the initial conditions do not lie on the conformal surface, one expects large 
coupling constant corrections to the anomalous dimensions, which will in turn
affect the strengths of the $\beta$-functions.

\section{Supergravity Duals}

The main support for the idea of a cascading RG flow in the original
case of the conifold 
comes from a supergravity dual construction. 
This dual reproduces the logarithmic decrease
in the effective number of colors towards the IR and also matches the
beta functions for the gauge couplings. 

In \cite{Franco:2004jz}, analog supergravity solutions were constructed describing
logarithmic cascades for the gauge theories on D-branes probing
complex cones over del Pezzo surfaces.
The fact that this was possible is remarkable, since they were
obtained without knowing the explicit
metric. These supergravity solutions are of the general
type studied by Gra\~{n}a and Polchinski \cite{Grana:2000jj}.

The general form of the metric is a warped product of flat
four-dimensional Minkowski space and a Calabi-Yau $\bX$
\beq
ds^2 = Z^{-1/2} \eta_{\mu\nu} dx^\mu dx^\nu + Z^{1/2} ds^2_\bX \ ,
\label{metric}
\eeq
The solution also carries 3-form flux $G_3 = F_3 - \frac{i}{g_s} H_3$.
In order to preserve $\mathcal{N}=1$ supersymmetry, $G_3$ must be
supported only on $X$, imaginary self-dual, a $(2,1)$ form and
harmonic. 
Indeed, it is possible to construct
a $G_3$ satisfying all these condition. It has the form
\beq
G_3 = \sum_{I=1}^n a^I (\eta + i \frac{dr}{r}) \wedge \phi_I 
\eeq
where the $\phi_I$, $I=1 \ldots n$, are a basis of $(1,1)$ forms
orthogonal to the 
K\"ahler class of the del Pezzo and $\eta = \left(\frac{1}{3} d\psi +
\sigma \right)$. 
The one-form $\sigma$ satisfies $d\sigma = 2 \omega$, with $\omega$
the K\"{a}hler form on $dP_n$, and $0 \leq \psi < 2\pi$ is the
angular coordinate on the circle bundle over $dP_n$.
 
The intersection product between the $\phi_I$ is 
$\int_{dP_n} \phi_I \wedge \phi_J = -A_{IJ}$, where $A_{IJ}$ 
is the Cartan matrix for the exceptional Lie algebra ${\mathcal E}_n$. 
There is a different type of fractional brane associated to each
$\phi_I$, given by D5-branes wrapping the 2-cycle in the del Pezzo
Poincar\'e dual to $\phi_I$.

Let us now study the number of D5-branes and D3-branes associated
to these solutions
\paragraph{D5-Branes:} 
The number of D5-branes is given by the Dirac quantization of the 
RR 3-form $F_3$:
$a^J = 6 \pi \alpha' M^J$.
Hence, this family of solutions are dual to cascades in which the
number of fractional branes of each type remains constant. 

\paragraph{D3-Branes:} Similarly, the effective number of D3-branes is
computed from $F_5 = {\mathcal F}_5 + *{\mathcal F}_5$ 
where
${\mathcal F}_5= d^4x \wedge d(Z^{-1})$ and $Z$ is the warp factor 
in \eref{metric}.
 The factor $Z$ satisfies the equation
\begin{equation}
\nabla^2_{\bf X} Z =- \frac{1}{6} |H_3|^2 \ .
\end{equation}
In \cite{Franco:2004jz}, $|F_3|^2$ was assumed to be
a function only of the radius, in which case
 \beq
Z(r) = \frac{2 \cdot 3^4}{9-n} \alpha'^2 g_s^2 
\left(\frac{\ln(r/r_0)}{r^4} + \frac{1}{4r^4} \right) 
	\sum_{i,j}  M^I A_{IJ} M^J
\eeq
and from Dirac quantization, the number of D3-branes
will grow logarithmically:
$N  = \frac{3}{2\pi} g_s \ln(r/r_0) \sum_{I,J} M^I A_{IJ} M^J$.
However, generically, $Z$ may depend on other coordinates
on the Calabi-Yau cone $\bf X$.\footnote{We would like to 
thank Q.~J.~Ejaz for telling us about this possibility.}  The function
$Z$ averaged over the other coordinates may still be logarithmic
in $r$ \cite{HK}.

\section{Recent developments}

Recently, there has been further progress in the study of quiver theories and their sugra duals. 
In \cite{Bertolini:2004xf}, a-maximization \cite{Intriligator:2003jj} was used to compute the volume 
of the 5d horizon of the dual of the $dP_1$ gauge theory, yielding an irrational value. This result 
corrected previous computations in the literature and was obtained by carefully taking into account the 
global symmetries that are actually preserved by the superpotential. 
In \cite{Franco:2004jz}, the duality cascade for $dP_1$ was
analyzed using
naive R-charges that did not take into account these global symmetries.
A stable elliptical region in coupling space was found with a
self-similar logarithmic cascade. Redoing the analysis with the new
R-charges, we find the same elliptical region albeit with a slightly
different shape and center.

The 5d horizon for the complex cone 
over $dP_1$ is called $Y^{2,1}$ and is a member of an infinite family of Sasaki-Einstein geometries denoted 
$Y^{p,q}$. They have $S^2 \times S^3$ topology. Their metrics were first found, locally, in
\cite{Gauntlett:2004zh} and then the global properties were analysed in
\cite{Gauntlett:2004yd}. In \cite{Martelli:2004wu}, their toric description was worked out. 
Furthermore, the gauge theory 
duals to the entire $Y^{p,q}$ family have been constructed 
\cite{Benvenuti:2004dy}. These developments change profoundly the status of the AdS/CFT, providing an infinite 
number of field theories with explicit sugra duals.

\begin{acknowledgments}
We would like to thank Qudsia Jabeen Ejaz, Ami Hanany, Pavlos Kazakopoulos,
Igor Klebanov and Joe Polchinski
for useful discussions. This
research is funded in part by the CTP and the LNS
of MIT and by the department of Physics at UPenn. 
The research of S.~F. was supported in part by
U.S.~DOE Grant $\#$DE-FC02-94ER40818.
The research of Y.-H.~H. was supported in part by
U.S.~DOE Grant $\#$DE-FG02-95ER40893 as well as
an NSF Focused Research Grant
DMS0139799 for ``The Geometry of Superstrings''.
C.~H. was supported in part by the NSF under
Grant No.~PHY99-07949. The research of J.~W. was supported by
U.S.~DOE Grant $\#$DE-FG02-90ER40542.
S.~F. would like to thank the organizers of Cargese Summer School,
where this material was presented.

\end{acknowledgments}



\begin{chapthebibliography}{99}


\bibitem{Maldacena:1997re}
J.~M.~Maldacena,
``The large N limit of superconformal field theories and supergravity,''
Adv.\ Theor.\ Math.\ Phys.\  {\bf 2}, 231 (1998)
[Int.\ J.\ Theor.\ Phys.\  {\bf 38}, 1113 (1999)]
[arXiv:hep-th/9711200].

\bibitem{Klebanov:2000hb}
I.~R.~Klebanov and M.~J.~Strassler,
``Supergravity and a confining gauge theory: Duality cascades and
chiSB-resolution of naked singularities,''
JHEP {\bf 0008}, 052 (2000)
[arXiv:hep-th/0007191].

\bibitem{Strassler}
M.~J.~Strassler,
``Duality in Supersymmetric Field Theory and an 
Application to Real Particle Physics,'' Talk given at International Workshop
on Perspectives of Strong Coupling Gauge Theories (SCGT 96), Nagoya,
Japan.  Available at 
http://www.eken.phys.nagoya-u.ac.jp/Scgt/proc/

\bibitem{Franco:2003ja}
S.~Franco, A.~Hanany, Y.~H.~He and P.~Kazakopoulos,
``Duality walls, duality trees and fractional branes,''
arXiv:hep-th/0306092.

\bibitem{Franco:2004jz}
S.~Franco, Y.~H.~He, C.~Herzog and J.~Walcher,
Phys.\ Rev.\ D {\bf 70}, 046006 (2004)
[arXiv:hep-th/0402120].

\bibitem{Grana:2000jj}
M.~Grana and J.~Polchinski,
``Supersymmetric three-form flux perturbations on AdS(5),''
Phys.\ Rev.\ D {\bf 63}, 026001 (2001)
[arXiv:hep-th/0009211].

\bibitem{HK}
C.~Herzog, Q.~J.~Ejaz,and I.~Klebanov,
``Cascading RG Flows for New Sasaki-Einstein Manifolds,''
arXiv:hep-th/0412193.

\bibitem{Bertolini:2004xf}
M.~Bertolini, F.~Bigazzi and A.~L.~Cotrone,
``New checks and subtleties for AdS/CFT and a-maximization,''
arXiv:hep-th/0411249.

\bibitem{Intriligator:2003jj}
K.~Intriligator and B.~Wecht,
``The exact superconformal R-symmetry maximizes a,''
Nucl.\ Phys.\ B {\bf 667}, 183 (2003)
[arXiv:hep-th/0304128].

\bibitem{Gauntlett:2004zh}
J.~P.~Gauntlett, D.~Martelli, J.~Sparks and D.~Waldram,
``Supersymmetric AdS(5) solutions of M-theory,''
Class.\ Quant.\ Grav.\  {\bf 21}, 4335 (2004)
[arXiv:hep-th/0402153].

\bibitem{Gauntlett:2004yd}
J.~P.~Gauntlett, D.~Martelli, J.~Sparks and D.~Waldram,
``Sasaki-Einstein metrics on S(2) x S(3),''
arXiv:hep-th/0403002.

\bibitem{Martelli:2004wu}
D.~Martelli and J.~Sparks,
``Toric geometry, Sasaki-Einstein manifolds and a new infinite class of
AdS/CFT duals,''
arXiv:hep-th/0411238, and references therein.

\bibitem{Benvenuti:2004dy}
S.~Benvenuti, S.~Franco, A.~Hanany, D.~Martelli and J.~Sparks,
``An infinite family of superconformal quiver gauge theories with
Sasaki-Einstein duals,''
arXiv:hep-th/0411264.


\end{chapthebibliography}
\end{document}